\documentclass[twocolumn,aps,showpacs]{revtex4}
\usepackage{epsfig}

\newcommand{\ba}{\begin{eqnarray}}
\newcommand{\ea}{\end{eqnarray}}

\begin{document}

\title{A study of randomness, correlations and collectivity in the nuclear shell model}

\author{V. Vel\'azquez, J. G. Hirsch, A. Frank}
\affiliation{Instituto de Ciencias Nucleares, 
Universidad Nacional Aut\'onoma de M\'exico, \\
Apartado Postal 70-543, 04510 M\'exico, D.F., M\'exico}
\author{A. P. Zuker} 
\affiliation{IRES, b\^at 27, IN2P3-CNRS
Universit\'e Louis Pasteur BP 28, \\
F-67037 Strasbourg Cedex 2, France}

\begin{abstract}
A variable combination of realistic and  random two-body interactions 
allows the study  of collective properties, such as the energy spectra
and  B(E2) transition strengths in  $^{44}$Ti, $^{48}$Cr
and $^{24}$Mg. It is found that the average energies of the yrast band states
maintain the ordering for any degree of randomness, but the B(E2) values 
lose their quadrupole collectivity when randomness dominates the Hamiltonian. 
The high probability of the yrast band to be ordered in the presence of 
pure random
forces exhibits the strong correlations between the different members
of the band.
\end{abstract}


\maketitle

\section{Introduction}

Nuclei display very regular spectral patterns. Low energy states in medium-
and heavy-mass even-even nuclei allow their classification in terms
of seniority, anharmonic vibrator and rotor nuclei \cite{zamfi94}, 
according to the ratio of the excitation energies of the states 
$4_1$ and $2_1$. While this regular behavior has been usually related
with specific forces, the investigation of the energy spectra with
random interactions \cite{joh98,joh99} has shown that
many body states are strongly correlated even in the presence of random
two-body interactions. Random interactions in bosonic Hilbert spaces, 
like those used in the IBM and the vibron model, exhibit a large predominance 
of vibrational and rotational spectra, strongly suggesting that in boson
spaces collectivity is an intrinsic property of the space of nuclear states
\cite{bij00,BF00_1,BF00_2}. Recently several studies were performed with 
random (tbre) and displaced random ensembles in order to simulate realistic 
systems \cite{hor01,vz02}, in particular to investigate the dominance of $0+$
states  \cite{zhao02_2,zhao02_3}

In the nuclear shell model the transition between random and collective
behavior in the energy spectra of $^{20}$Ne generated by two-body forces 
was addressed in \cite{cor82}. 
Collectivity was generated with a quadrupole-quadrupole
force, while a residual random interaction was included in the Hamiltonian 
in order to study its consequences on the system's spectroscopic 
properties. Both the eigenvalue distribution and the overlap between the 
SU(3) and calculated wave functions exhibit the smooth path from a 
Hamiltonian
dominated by the collective force to a random, non-collective, one.

The present work aims to extend these ideas by studying the transition from
a realistic parametrization of the two-body force to a purely random one, 
in complex systems like $^{24}$Mg, $^{44}$Ti and $^{48}$Cr. The realistic
interactions we have chosen are the universal Wildenthal \cite{Wil} 
interaction for the sd-shell,
and the KB3 \cite{kb3} for the fp-shell. The probability for each state 
in the yrast band
to follow a sequence where the higher energies correspond to the states with 
the larger angular momentum is studied by varying the mixing between realistic 
and random forces in the Hamiltonian. The evolution of the average energies for
each member of the band, as well as its B(E2) values, is also reported.

\section{Band structure}

The combination of random and realistic interactions is taken as ~\cite{cor82}
\ba
H = a H_C + b H_R  \hbox{,~~~with~~~}
a + b = 1 ,
\ea
where $H_C$ is a realistic Hamiltonian 
and  $H_R$ is a two body random ensemble.
Both $H_C$ and $H_R$ are written as \cite{DZ}
\begin{equation}
H=\sum_{r \le s, t \le u} V_{rstu}^{JT} Z^\dagger_{rs,JT}\cdot Z_{tu,JT}~,
\end{equation}
in terms of scalar products of the normalized pair creation operators
$Z^\dagger_{ij,JT}$ and its Hermitian conjugate $Z_{kl,JT}$, 
where $r,s,\ldots$ specify sub-shells associated with individual orbits,
and $J, T$ the coupled angular momentum and isospin. 
For the  realistic Hamiltonian  $V_{rstu}^{JT}$ is the Wildenthal or $KB3$ interaction.
For the random case the $V_{rstu}^{JT}$ matrix elements are taken from a
two body  random ensemble (TBRE), i.e., to be real and normally
distributed with mean zero and width $\sigma$ for the off-diagonals
and $\sqrt{2} \sigma$ for the diagonals. 
The values of the width $\sigma$ are taken from the realistic interactions:
1.34 MeV and 0.60 MeV, for the sd- and fp-shells, respectively.
The parameter $b$ is varied from 0 to 1, to cover the different mixing
from the realistic interaction to a pure random force.

\subsection{The fp shell}

As a first example we take $^{48}$Cr, a rotational nucleus which
has been widely studied in full shell model calculations.
In the present case the diagonalization is performed in the 
$f_{\frac{7}{2}} p_{\frac{3}{2}}$ shell with the code Antoine \cite{cau89}, 
using a KB3 interaction \cite{pz90} without single particle energies. 
Fig. 1 shows the average energies $\overline{E_J}$ of the lowest energy
state for each angular momentum J= 0, 2, ..., 16 $\hbar$, the yrast band, 
calculated for 960 samples of the random interaction.

\begin{figure}[h]  
  \begin{center}
    \leavevmode 
    \psfig{file=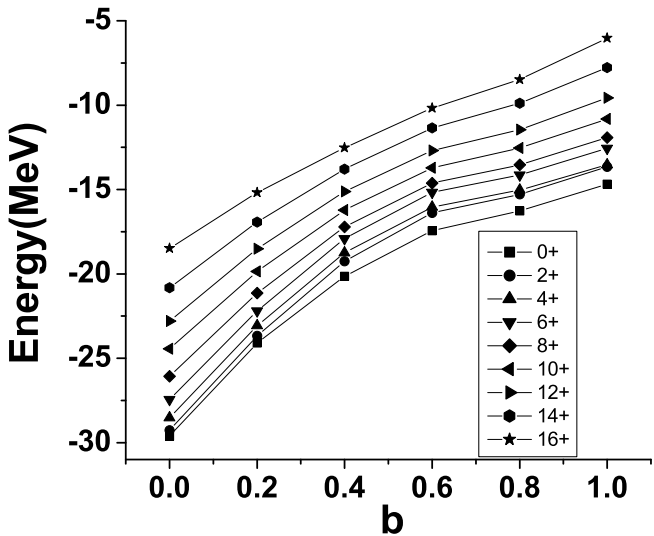}
    \caption{Average energy of 960 runs for each angular momentum 
of the ground band in $^{48}$Cr. }
    \label{fig1}
  \end{center}
\end{figure}

 The vertical axis displays
the average energy of each state, as a function of the mixing parameter
$b$. At the left hand side the realistic energies are shown, with
its distinctive rotor pattern. As $b$ increases to the right, the
order between the different members of the bands is maintained, but
their relative separation changes.  The average 
energies in the right hand side, the pure random Hamiltonian, still exhibit a band 
structure but have lost their quadrupole collectivity, as discussed below in connection with their 
B(E2) values.

The evolution of the average ground state band of $^{44}$Ti, calculated in the
full pf-shell, is shown in Fig. 2 as a function of the mixing parameter
$b$. It can be seen that the relative ordering of the states
with different angular momentum is maintained, but their relative
separations vary significantly in the transition from the realistic to the
random Hamiltonian. However, up to $b = 0.6$ the change is mostly a scale 
variation, with the nearly equidistant structure of the band keeping its form.
The energy spectra evolves from a vibrational equidistant form at $b=0$
to a mixed, yet ordered, spectrum for pure random forces.
The average ground state energy increases with the mixing to a maximum value, 
with a slight decrease for $b = 1$.

\begin{figure}[htb]  
  \begin{center}
    \leavevmode 
    \psfig{file=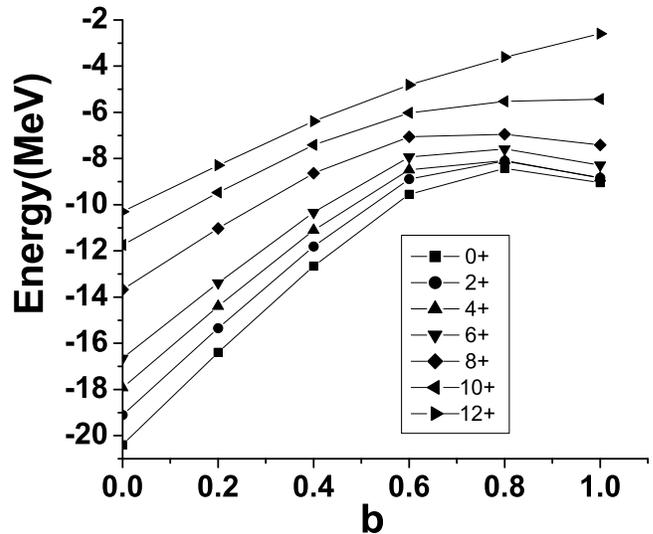}
    \caption{Average energy of 960 runs for each angular momentum 
of the ground band in $^{44}$Ti.}
    \label{fig2}
  \end{center}
\end{figure}

Similar patterns of evolution of the energy centroids for each angular
momentum, as function of the mixing parameter, were found in several
other nuclei like $^{46,48}$Ca and $^{46}$Ti. While for $^{46}$Ti,
the centroid of the J=2 state is lower in energy than the J=0 state
for a pure random Hamiltonian, in general for these nuclei and others
in the sd-shell the average energy ordering is conserved, and there is a 
gradual change in the energy spectra. 

The width $\sigma$ associated with the lowest average energy for each angular momentum, 
calculated as the square root of the variance, is shown in Fig. 3 for
the gs-band of $^{48}$Cr, as a function of the angular momentum,   
for different values of the mixing parameter $b$.
The energy width increases with the mixing, from 1.5 MeV for $b=0.2$
to 7-9 MeV for $b=1.0$. Given that, as $b$ increases, these widths become larger
than the gaps between the average energies, they indicate a strong mixing
between different bands, an effect closely associated with the lack of collectivity
discussed bellow. As a function of the angular momentum all the widths are
essentially flat, showing a moderate increase for the largest angular momenta.

\begin{figure}[h!]  
  \begin{center}
    \leavevmode 
    \psfig{file=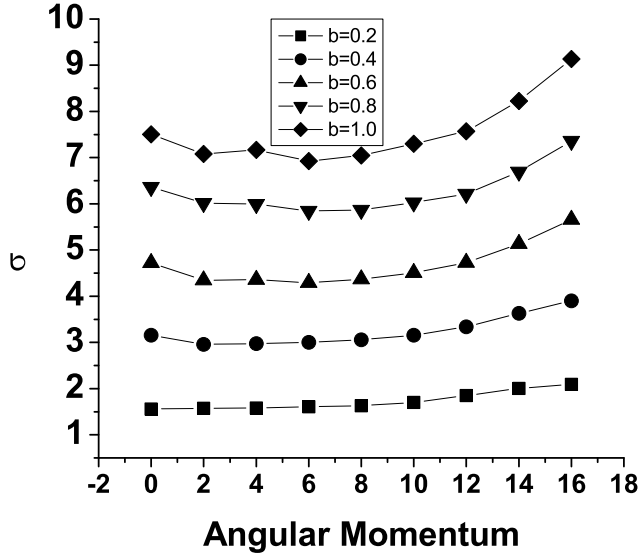}
    \caption{Average energy width of 960 runs for each angular momentum 
of the ground band in $^{48}$Cr. }
    \label{width}
  \end{center}
\end{figure}

The energy widths are larger than the relative energy differences 
between adjacent states in the g.s. band, implying that there are
individual cases in which the spectrum is not ordered. However, the
number of runs with a completely ordered spectrum is large even for
pure random interactions, as is discussed in detail bellow.

\subsection{The sd-shell}

Calculations in the sd-shell can be done for the full  
shell and an arbitrary number of active particles.  The realistic interaction 
used in this case is the universal Wildenthal interaction (USD \cite{Wil}). 
The nucleus $^{24}$Mg offers a rich enough system, where three bands can be 
studied simultaneously. They correspond, for $b = 0$, to the ground state (gs),
$\beta$- and $\gamma$- bands. The first two start with J=0 and contain only
even-J states, while the $\gamma$- band starts with J = 2 and includes states 
with both even and odd angular momenta, up to J = 8. 
Following the evolution of the average
energy levels as the mixing parameter $b$ increases, we reconstruct the three
bands for each value of $b$. Nevertheless, for a pure random interaction
one can show that the states have lost their quadrupole collectivity (see 
bellow), 
and for this reason they do not form bands in the usual sense.

In Fig. 4 the average energies for each angular momentum, for
the ground, $\beta$- and $\gamma$- bands in $^{24}$Mg, are presented
for different values of the mixing parameter $b$.

\begin{figure}[h!]  
  \begin{center}
    \leavevmode 
    \psfig{file=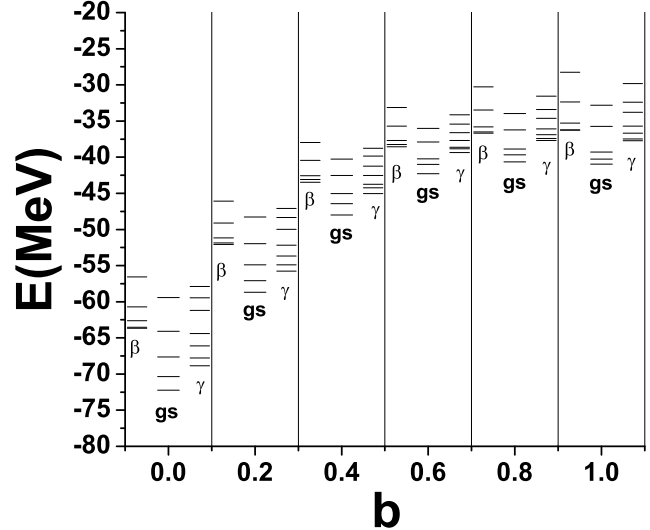,width=9.cm}
    \caption{Average energy of 960 runs for each angular momentum 
of three bands in $^{24}Mg$, see text for labels description.}
    \label{fig4}
  \end{center}
\end{figure}

The most relevant features observed in Fig. 4 are that the three bands
evolve in a similar way, keeping their internal order as well as the
relative separation between the bands.
The correlation between increasing
energy and angular momentum is strictly followed in all
the averaged bands. The absolute ground state energy becomes larger
as the mixing parameter increases, following the same pattern found
in the fp-shell.

\section{Ordering probabilities}

In order to further study these systems, we have analyzed the probability
that each state in the band has  the usual ordering i.e, the 
largest the
angular momentum, the highest the energy for a given band. To do so 
we counted the number of runs where the first J=0 state is the ground state,
or the first excited state in the band,  the second excited state, and so on, 
in the 960 runs. We did the same for the  states with angular momenta J = 2, 
4, ... In Table I the percentages  for $^{48}$Cr with a mixing of  
$ b=0.4$ are listed. All the states have a probability of at least 
 90\% to occupy its physically expected place, while the dispersion is very small.

\begin{table}[h]
\begin{center}
\begin{tabular}{cccccccccc}
$J+$  &1st & 2nd & 3rd& 4th & 5th & 6th & 7th& 8th & 9th \\
\hline

$0$  & 96.56&  1.04&  1.25&  0.52&  0.42&  0.10&  0.00&  0.10&  0.00\\
$2$  &  1.56& 91.88&  5.31&  0.62&  0.21&  0.21&  0.10&  0.00&  0.10\\
$4$  &  0.21&  5.62& 92.29&  1.04&  0.42&  0.21&  0.21&  0.00&  0.00\\
$6$  &  0.52&  0.94&  0.42& 92.50&  4.27&  0.94&  0.21&  0.00&  0.21\\
$8$  &  0.73&  0.21&  0.21&  3.75& 93.02&  1.04&  0.31&  0.52&  0.21\\
$10$ &  0.00&  0.00&  0.10&  0.83&  0.52& 95.21&  2.29&  0.83&  0.21\\
$12$ &  0.00&  0.00&  0.21&  0.62&  0.83&  2.08& 96.04&  0.21&  0.00\\
$14$ &  0.21&  0.21&  0.10&  0.10&  0.31&  0.10&  0.52& 97.40&  1.04\\
$16$ &  0.21&  0.10&  0.10&  0.00&  0.00&  0.10&  0.31&  0.94& 98.23

\end{tabular}

\caption{ Probability for each angular momentum state to be in the
correct position in the ground state band with $b=0.4$ for $^{48}$Cr.} 
\label{tab1}
\end{center}
\end{table}

Table II lists the probability for states of each angular momentum in 
$^{48}$Cr to occupy the indicated place for a purely random Hamiltonian,
i.e. $b = 1$. The probabilities of being in its expected place run 
from 44\% for J = 2 to 83\% for
J = 16. States with J = 0, 2 and 4 are those which more often
fail to occupy their place, as they tend to exchange positions. These 
results represent an extension of
previous studies concerning the probability of each state to be the ground 
state, listed in the first column \cite{Zhao01}.

\begin{table}[h]
\begin{center}
\begin{tabular}{cccccccccc}
$J+$  &1st & 2nd & 3rd& 4th & 5th & 6th & 7th& 8th & 9th \\
\hline

$0$ & 60.31&  8.85&  9.27&  4.48&  4.90&  2.50&  1.98&  2.29&  5.42\\
$2$ & 17.40& 43.54& 16.98&  4.17&  3.96&  2.60&  3.12&  5.00&  3.23\\
$4$ &  6.56& 27.71& 47.08&  6.46&  2.40&  2.71&  4.38&  2.19&  0.52\\
$6$ &  2.40&  5.42&  9.79& 64.58&  6.25&  5.31&  3.12&  1.77&  1.35\\
$8$ &  3.12&  3.65&  4.69&  9.06& 69.58&  6.46&  1.15&  1.56&  0.73\\
$10$ &  1.35&  2.50&  3.44&  5.94&  7.19& 74.69&  3.54&  0.62&  0.73\\
$12$ &  1.98&  1.88&  5.52&  2.50&  3.23&  3.65& 79.06&  1.98&  0.21\\
$14$ &  1.15&  5.62&  1.98&  1.67&  0.94&  1.25&  2.40& 80.10&  4.90\\
$16$ &  5.73&  0.83&  1.25&  1.15&  1.56&  0.83&  1.25&  4.48& 82.92

\end{tabular}

\caption{ Probability for each angular momentum state to be in different 
positions, for  the $^{48}$Cr ground state band with $b=1.0$.} 
\label{tab2}
\end{center}
\end{table}

Table III displays the probability for each angular momentum state to be in
a given position, in the $^{44}$Ti ground state band with $b=1.0$.
These probabilities have values between 44\% for J=2 to 95\% for J = 12.
The fact that the states with J=2 and 4 have less than 50\% probability
of occupying in their places is strongly connected with the closeness
of their average energies, shown in Fig. 2 for $b = 1$.

\begin{table}[h]
\begin{center}
\begin{tabular}{ccccccccc}
$J+$   &1st & 2nd & 3rd& 4th & 5th & 6th & 7th  \\
\hline
$0$ & 46.25& 19.27& 20.42&  6.98&  5.62&  1.35&  0.10\\
$2$ & 16.56& 39.90& 24.38& 11.56&  5.83&  1.25&  0.52\\
$4$ & 21.04& 25.83& 40.10&  9.38&  3.02&  0.21&  0.42\\
$6$ &  8.44& 10.42& 10.10& 64.58&  5.42&  1.04&  0.00\\
$8$ &  6.56&  3.96&  4.27&  5.83& 76.77&  2.08&  0.52\\
$10$ &  0.10&  0.52&  0.52&  1.25&  2.60& 92.08&  2.92\\
$12$ &  1.04&  0.10&  0.21&  0.42&  0.73&  1.98& 95.52
\end{tabular}

\caption{ Probability for each angular momentum to be in 
different positions, for the $^{44}$Ti ground state band, with $b=1.0$.} 
\label{tab3}
\end{center}
\end{table}

Tables IV, V and VI display the probability that states with different angular 
momenta in $^{24}$Mg have to occupy a given place in each band for $b=1.0$,
for the ground state, $\beta$- and $\gamma$- band, respectively. 
In most cases the diagonal probability, i.e. the probability that each state occupies
its expected place, is larger than 50\%. The exceptions are the states with J = 2 and 3
in the $\gamma$-band, and those with J= 0 and 2
in the $\beta$-band, whose probabilities lie between 38\% and 48\%.

\begin{table}[h]
\begin{center}
\begin{tabular}{cccccc}
$J+$  &1st & 2nd & 3rd& 4th & 5th   \\
\hline

$0$ & 56.56& 12.40& 12.29&  6.77& 11.98\\
$2$ & 18.12& 52.81& 13.12&  9.79&  6.15\\
$4$ &  9.06& 19.58& 63.85&  5.52&  1.98\\
$6$ &  4.38& 11.25&  6.98& 71.25&  6.15\\
$8$ & 11.88&  3.96&  3.75&  6.67& 73.75

\end{tabular}

\caption{ Probability for states belonging to the ground state band 
in $^{24}$Mg, with $b=1.0$, to occupy different positions.}
\label{tab4}
\end{center}
\end{table}

\begin{table}[h]
\begin{center}
\begin{tabular}{cccccccc}
$J+$  &1st & 2nd & 3rd& 4th & 5th & 6th & 7th \\
\hline

$2$ & 45.21& 22.92&  7.50&  5.94&  4.79&  6.46&  7.19\\
$3$ & 27.40& 38.33& 13.54&  6.15&  4.17&  6.04&  4.38\\
$4$ &  7.40& 14.90& 50.62& 15.94&  7.29&  3.02&  0.83\\
$5$ &  7.71& 12.08& 15.10& 58.85&  5.00&  1.04&  0.21\\
$6$ &  3.44&  3.75&  6.25&  7.60& 66.25& 11.88&  0.83\\
$7$ &  5.10&  4.27&  4.17&  3.96& 10.83& 64.38&  7.29\\
$8$ &  3.75&  3.75&  2.81&  1.56&  1.67&  7.19& 79.27

\end{tabular}

\caption{ The same as in Table 4, for the $\gamma$-band.}
\label{tab5}
\end{center}
\end{table}

\begin{table}[h]
\begin{center}
\begin{tabular}{cccccc}
$J+$  &1st & 2nd & 3rd& 4th & 5th   \\
\hline

$0$ & 46.67& 20.73& 13.54&  8.44& 10.62\\
$2$ & 28.02& 47.40& 12.81&  8.44&  3.33\\
$4$ & 13.23& 19.27& 65.62&  1.88&  0.00\\
$6$ &  6.35&  9.38&  5.94& 77.50&  0.00\\
$8$ &  5.73&  3.23&  2.08&  3.75& 85.21

\end{tabular}

\caption{ The same as in Table 4, for the $\beta$-band.} 
\label{tab6}
\end{center}
\end{table}

\section{Collectivity}

A sensitive measure of the quadrupole collectivity of the system is the 
B(E2,$2_1 \rightarrow 0_{gs}$) transition strength.  The distribution of B(E2)
strengths in $^{48}$Cr, for four different values of the mixing 
parameter $b$, is shown in Fig. 5. Its experimental value is 230 $e^2b^2$.
For $ b=0.2$, shown in insert a), the distribution is concentrated around
the experimental B(E2) value. 
For $ b=0.4$, insert b), the increase in the random
components of the Hamiltonian leads to an important 
fragmentation of the B(E2) intensity with four clusters, one near zero, 
the second one around  75 $e^2b^2$, the next one
near 170 $e^2b^2$ and the last one close to the measured value.
For $b = 0.6$, insert c), most of the B(E2) values are very
small, with some intensity at the collective B(E2) values.
Finally, in insert d) the distribution of B(E2) values for a purely random
Hamiltonian is shown. It is strongly concentrated at very small values, showing
a complete lack of quadrupole collectivity, in consonance with the findings 
of Ref. \cite{cor82}.

\begin{figure}[h]  
  \begin{center}
    \leavevmode 
    \psfig{file=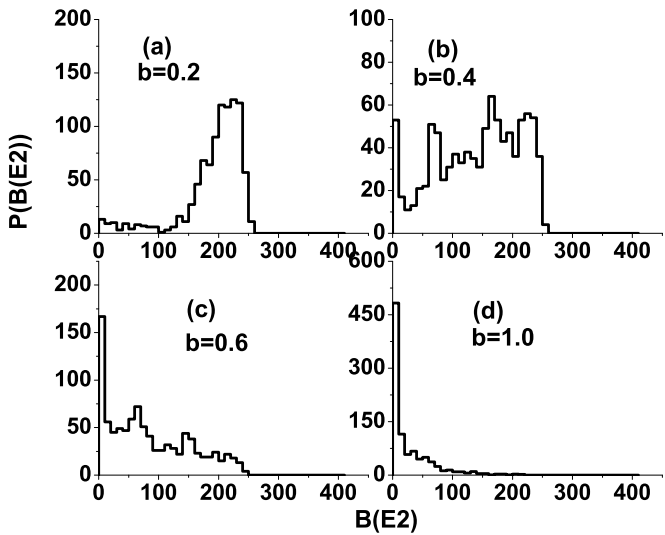}
      \caption{Probability densities for B(E2,$2\rightarrow$ 0) for 
$^{48}$Cr.}
      \label{fig5}
    \end{center}
\end{figure}

Fig. 6 displays the distribution of B(E2,$2\rightarrow$ 0) values 
for $^{44}$Ti. 
Insert a) shows the results for $b=0.2$, with a
narrow distribution around the measured value of 147 $e^2b^2$.
In insert b) the distribution for $b=0.4$ is presented, which is concentrated 
around the same value. For $b=0.6$, Fig. 5 c), the distribution
is still concentrated around the collective B(E2) values despite the dominance
of the random component in the Hamiltonian. However, this collectivity is 
completely lost when pure random forces are employed, as shown in Fig. 5 d).

\begin{figure}[h]  
  \begin{center}
    \leavevmode 
    \psfig{file=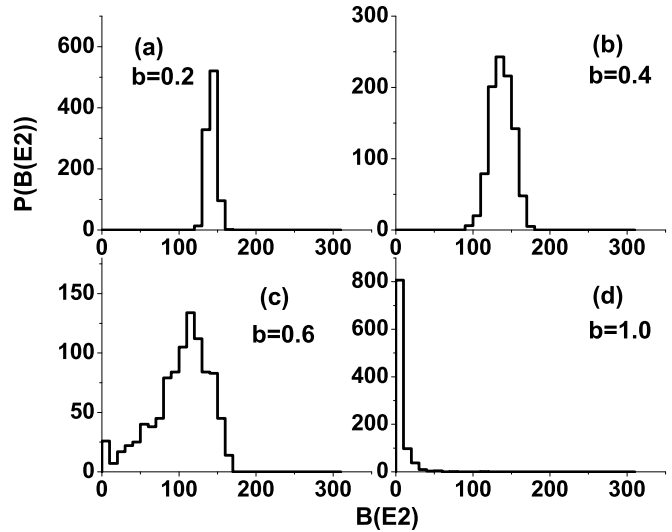}
    \caption{ Probability density for B(E2,$2\rightarrow$ 0)for 
$^{44}$Ti. }
    \label{fig6}
  \end{center}
\end{figure}

A useful indicator of collectivity is the energy ratio
\ba
R = {\frac {E_{4^+_1}} {E_{2^+_1}}  }.
\ea
The value R = 2 is associated with a harmonic oscillator spectrum,
while R = 3.3 characterizes a rigid rotor structure. 

Fig. 7 shows the distribution of energy ratios for $^{48}$Cr.
The case  $b = 0.2$, shown in insert a), does exhibit the actual rotor 
behavior of this nucleus. This feature remains dominant for $b = 0.4$,
insert b), while for $b = 0.6$ the distribution of energy ratios is
wide and peaked at R=1. For $b = 1.0$, shown in Fig 7 d), the distribution 
is very wide, with a clear dominance of the R=1 ratio, in correspondence
with the near degeneracy of the states with J= 2 and 4 for $b =1$, as
shown in Fig. 1.

\begin{figure}[h]  
  \begin{center}
    \leavevmode 
    \psfig{file=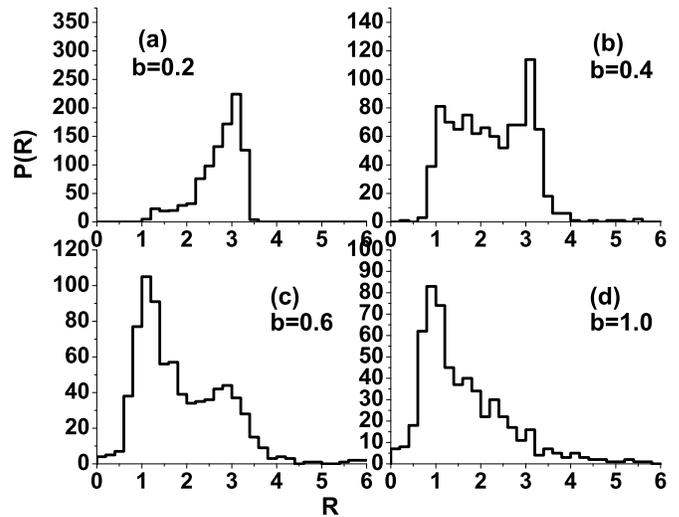}
      \caption{Probability densities for the energy ratios R for 
$^{48}$Cr.}
      \label{fig7}
    \end{center}
\end{figure}

$^{44}$Ti has a structure closer to a harmonic oscillator than to a rotor. 
This feature can be seen in Fig. 8 a), which displays the distribution
of energy ratios for $b = 0.2$, narrowly concentrated around R = 2.
For $b = 0.4$ and $0.6$, inserts b) and c), the vibrational structure 
is wider but well defined. For a pure random interaction, Fig. 
8 d), the distribution  is peeked at R = 1, reflecting the near degeneracy
of the average energies $\bar E_2$ and $\bar E_4$.
The displacement of the most probable energy ratio R to 1 is
accompanied by a lack of quadrupole coherence, in consistent
with the previous analysis of the B(E2) transition strengths.

\begin{figure}[h]  
  \begin{center}
    \leavevmode 
    \psfig{file=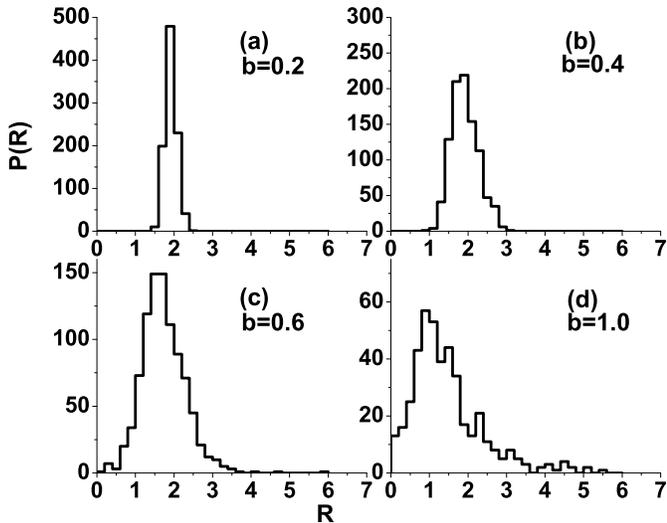}
    \caption{ Probability densities for energy ratios R for 
$^{44}$Ti. }
    \label{fig8}
  \end{center}
\end{figure}

\section{Correlation and coherence.}

Having discussed the evolution of the energy centroids as a function of the
mixing parameter $b$, and the probability for each state
to occupy a certain position in the band, it is natural to
study  the probability that the ground state band has all its 
states properly ordered. This probability, given as the percentage of
results from the 960 runs which are properly ordered, is shown in Fig. 9 for
$^{48}$Cr, $^{44}$Ti and $^{24}$Mg. In the three cases it is apparent 
that the spectrum is always ordered when the realistic interaction
dominates the Hamiltonian, and that the probability for the ground
state band to be ordered decreases to about 35\% for a fully random 
Hamiltonian.

\begin{figure}[h]  
  \begin{center}
    \psfig{file=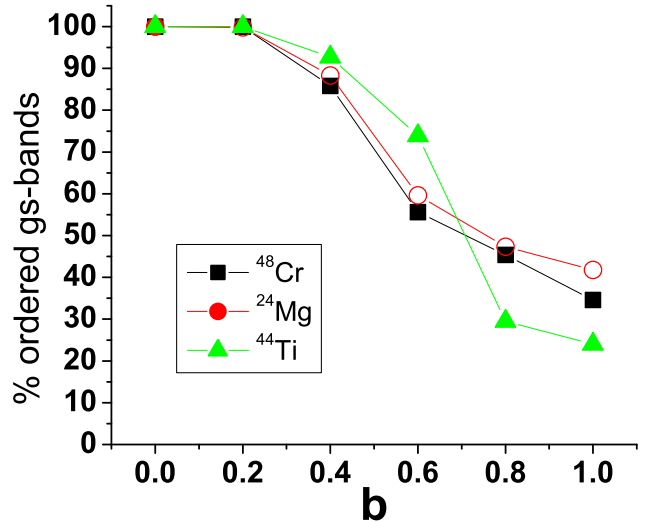}
    \caption{Percentage of ordered bands vs $b$ for Cr,Ti and Mg.}
    \label{fig9}
  \end{center}
\end{figure}

It is worth emphasizing that
 if these probabilities were independent of each other, the probability 
to find
a completely order band would be the product of the probabilities for each 
state,
i.e., the product of the diagonal elements of the matrices shown in Tables 
II, III and IV. However, these
products are all smaller than 0.04, far smaller than the probabilities of 25 -
45 \% found in these calculations. These results show that the
many body states obtained with two-body random forces are strongly correlated.
This correlation in their energies is not, however, associated with 
quadrupole coherence, which is lost for pure random forces \cite{cor82}. The 
subtle connection between correlation and coherence requires further analysis.

The {\em correlation} $r$ for each value of the mixing parameter $b$,
between the states with energies $E_J$ and $E_{J+2}$ in the sample, 
can be quantitatively calculated as \cite{DP}
    
\ba
r = {\frac {cov(E_J,E_{J+2})} {\sigma_{E_J}\sigma_{E_{J+2}}}} 
\ea
where
\ba
cov(E_J,E_{J+2})= \frac{\sum_i^n ({E_J}_i- {\overline{E_J}})({E_{(J+2)}}_i- 
{\overline{E_{J+2}}})}{n}
\ea
Here, ${cov(E_J,E_{J+2})}$ is the covariance of the two distributions,
$\sigma $ is the square root of the variance $cov(E_J,E_J)$, and $n$ is 
the size of the space.  
If $r=1$ the distributions are fully correlated, while if $r \approx 0$ 
there is no correlation between them. 
In table VII the correlation $r$ for pairs of energies in the ground state 
band of
$^{48}$Cr are listed .  
\begin{table}[h]
\begin{center}
\begin{tabular}{cccccc}
$J,J+2$  &0.2 & 0.4 & 0.6& 0.8 & 1.0   \\
\hline

$0+,2+$     &0.966&     0.928&      0.937&      0.949&     0.991\\
$2+,4+$     &0.996&     0.991&      0.991&      0.991&     0.986\\
$4+,6+$     &0.992&     0.977&      0.974&      0.981&     0.980\\ 
$6+,8+$     &0.991&     0.989&      0.992&      0.993&     0.993\\
$8+,10+$    &0.987&     0.989&      0.989&      0.989&     0.987\\
$10+,12+$   &0.985&     0.982&      0.982&      0.984&     0.985\\
$12+,14+$   &0.992&     0.989&      0.987&      0.986&     0.986\\
$14+,16+$   &0.992&     0.992&      0.990&      0.989&     0.988

\end{tabular}

\caption{ Correlations $r$ between neighbor states in ${48}Cr$.} 
\label{tab7}
\end{center}
\end{table}

The presence of strong correlations between the different wave
functions obtained with two-body random forces, suggests that the 
many-body states could be well approximated
by a small number of configurations which may correspond to definite 
shapes, as was
found for bosonic models. This would imply that the very large number 
of shell-model
many-body states would be limited or constrained by the geometry imposed by the
existence of a two-body Hamiltonian, even for the case that its components
are randomly selected \cite{chau}.

\section{ The DTBRE case}

In \cite{vz02} it was shown that a displaced two body random ensemble (DTBRE)
gives rise to coherent rotor patterns. To complement the present
study we analyze in this section the transition from the realistic KB3
interaction to a DTBRE in the ground state band of $^{48}$Cr.
The DTBRE correspond to matrix elements with a normal distribution 
centered at $c=-1.0$ and
width $\sigma=0.6$. Fig. 10 shows the evolution of the average 
energies in the ground-state band of $^{48}$Cr as a function of the mixing parameter $b$. 
\begin{figure}[h]  
    \leavevmode 
    \psfig{file=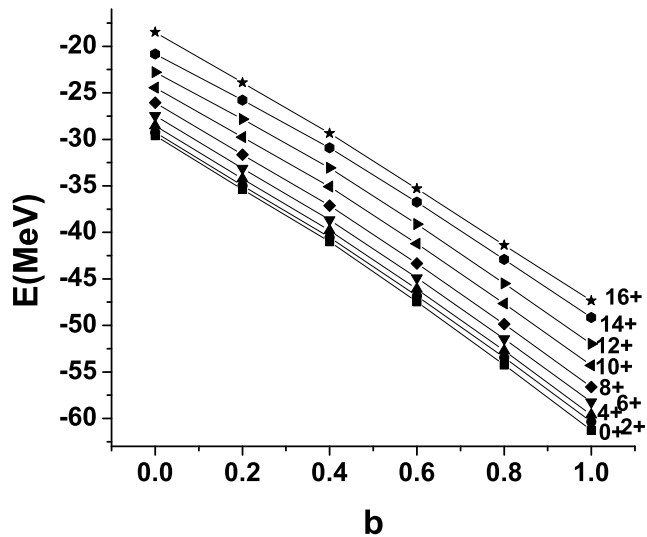, width=9.cm}
    \caption{ Average energy of 960 runs for each angular momentum 
of the ground band in $^{48}$Cr with a DTBRE mixing. }
    \label{fig10}
\end{figure}

At variance from the results shown in Fig. 1, 2 and 4, when the
random component of the interaction increases the absolute energies
continuously decrease. As expected, the rotor structure survives
the transition from realistic to random interactions with nearly no changes. 

In Table VIII we show the probability for states belonging to the ground 
state band in $^{48}$Cr, to occupy its expected place. Notice that even for 
$b=1.0$, when the pure displaced random ensemble is employed, the probability 
for each state to be in its place is larger than $88\%$,
thus exhibiting a high ordering.
The percentage of fully ordered bands for each mixing is equally large:
100\% for $b=0.2$, 89\% for $b=0.6$ and 70\% for $b=1.0$.
If the centroid of the DTBRE is displaced to more negative values, like $c=-3.0$,
100 \% of the bands are ordered \cite{vz02}. The distribution of B(E2) probabilities
exhibits a clear presence of quadrupole coherence for $c = -1$. 
They are concentrated in a narrow peak around the collective B(E2) values
for $c = -3$ \cite{vz02}.

\begin{table}[h]
\begin{center}
\begin{tabular}{cccccc}
$J+$  &$b=0.2$ & $0.4$ & $0.6$& $0.8$ & $1.0$   \\
\hline

$0$ & 99.58& 99.79& 97.71& 92.08& 88.12\\
$2$ & 98.75& 99.27& 95.00& 86.04& 80.31\\
$4$ & 98.96& 99.38& 95.21& 86.46& 79.90\\
$6$ & 99.27& 99.58& 96.35& 91.46& 87.50\\
$8$ & 99.27& 99.69& 98.02& 94.69& 91.15\\
$10$ & 99.58& 99.90& 98.96& 96.77& 94.38\\
$12$ & 99.69& 99.90& 98.65& 96.77& 95.00\\
$14$ & 99.58& 99.58& 94.79& 90.62& 87.92\\
$16$ & 99.58& 99.69& 95.00& 90.42& 88.33

\end{tabular}

\caption{ Probabilities for states belonging to the ground state band 
in $^{48}$Cr, to occupy their ordered place, for a DTBRE mixing.}
\label{tab8}
\end{center}
\end{table}
  
\section{Summary and conclusions}

The average energies of states with different angular momentum preserve 
their ordering inside the band when the Hamiltonian is changed smoothly
from a realistic to a random one. Ground state energies increase as 
a function of the mixing parameter. 
The quadrupole collectivity is lost when the Hamiltonian
is dominated by random two-body forces, and the probability that the
ground state band remains ordered diminishes to 25-45\% in the random limit,
which is anyway far larger than the product of the probabilities for
each state to be in its place, thus exhibiting the strong correlations between
the different wave functions.
On the other hand, when displaced two-body random ensembles are
employed, the average energies decrease as the random component increases,
and the rotor pattern remains unchanged.

\acknowledgments

  The exact diagonalizations were performed with the ANTOINE code.
This work was supported in part by Conacyt, M\'exico.

\end{document}